\documentclass[preprint,showpacs,amsmath,amssymb]{revtex4}
\usepackage{amssymb}
\usepackage{graphicx}
\usepackage{subfig}
\usepackage{tabularx}
\usepackage{dcolumn}% Align table columns on decimal point
\usepackage{bm}% bold math
\usepackage{multirow} %for table
\begin{document}

\title{Analysis of the Crystal Ball data on $K^-p\to\pi^0\Sigma^0$ reaction with center-of-mass energies of $1536\sim 1676$ MeV }

\author{Jun Shi$^{1}$}
\author{Bing-Song Zou$^{2,1}$}

\affiliation{$^1$ Institute of High Energy Physics and Theoretical
Physics Center for Sciences Facilities, Chinese Academy of Sciences,
Beijing 100049, China\\
$^2$ State Key Laboratory of Theoretical Physics, Kavli Institute
for Theoretical Physics China, Institute of Theoretical Physics,
Chinese Academy of Sciences, Beijing 100190,China}
%\date{\today}

\begin{abstract}
With an effective Lagrangian approach, we analyze the $K^-p\to
\pi^0\Sigma^0$ reaction to study the $\Lambda$ hyperon resonances by
fitting the Crystal Ball data on differential cross sections and
$\Sigma^0$ polarization with the center-of-mass energies of
$1536\sim 1676$ MeV. Besides well established PDG 4-star $\Lambda$
resonances around this energy range, the $\Lambda(1600){1\over 2}^+$
resonance, listed as a 3-star resonance in PDG, is found to be
definitely needed. In addition, there is strong evidence for the
existence of a new $\Lambda({3\over 2}^+)$ resonance around 1680
MeV.

\end{abstract}
\pacs {13.75.-n, 13.75.Cs, 14.20.Gk, 25.75.Dw}
\maketitle{}

\section{INTRODUCTION}
The $\overline{K}N$ scattering interaction has been widely used to
study the hyperon resonances. In our previous work
~\cite{pzgao,Sigma}, we have analyzed the $K^-N\rightarrow
\pi\Lambda$ to study the $\Sigma$ resonances, now we move forward to
study the pure isospin-0 reaction $K^-p\rightarrow\pi^0\Sigma^0$ to
learn structures of the $\Lambda$ resonances.

Many studies have been carried out to investigate the $\Lambda$
resonances. Oset et al.~\cite{Oset1,Oset2} used a chiral unitary
approach for the meson-baryon interactions and got two $J^P={1\over
2}^-$ resonances with one mass near 1390 MeV and the other around
1420 MeV. They believe the well established $\Lambda(1405){1\over
2}^-$ resonance listed in PDG~\cite{PDG} is actually a superposition
of these two ${1\over 2}^-$ resonances. Manley et al.~\cite{Manley}
and Kamano et al.~\cite{Kamano} made multichannel partial-wave
analysis of $\overline{K}N$ reactions and got results with some significant
differences. Zhong et al.~\cite{xhZhong} analyzed the
$K^-p\rightarrow\pi^0\Sigma^0$ reaction with the chiral-quark model
and discussed characteristics of the well established $\Lambda$
resonances. Liu et al.~\cite{Xie_eta} analyzed the
$K^-p\rightarrow\eta\Lambda$ reaction with an effective Lagrangian
approach and implied a D03 resonance with mass about 1670 MeV but
much smaller width compared with the well established
$\Lambda(1690){3\over 2}^-$. So there are still some ambiguities of
the $\Lambda$ resonant structures needing to be clarified.

Recently, the most precise data on the differential cross sections
for the $K^-p\to\pi^0\Sigma^0$ reaction have been provided by the
Crystal Ball experiment at AGS/BNL~\cite{CB08,CB09}. The $\Sigma^0$
polarization data were presented for the first time. However, with
different data selection cuts and reconstructions, two groups in the
same collaboration, {\sl i.e.}, VA group~\cite{CB08} and UCLA
group~\cite{CB09}, got inconsistent results for the $\Sigma^0$
polarizations. Previous multi-channel
analysis-\cite{Manley,Kamano,xhZhong} of the $\overline{K}N$ reactions failed
to reproduce either set of the polarization data.

In the present work, instead of performing some sophisticated
multi-channel analysis to stuck into various problems, as the first
step, we concentrate on the most precise data by the Crystal Ball
collaboration on the pure isospin scalar channel of $\overline{K}N$ reaction to
see what are the $\Lambda$ resonances the data demand and how the
two groups' distinct polarization data~\cite{CB08,CB09} influence
the spectroscopy of $\Lambda$ resonances. Consistent differential
cross sections of earlier work by Armenteros et
al.~\cite{LowEnergyData} at lower energies are also used.

This work is organized as follows. In Sec. \ref{theory} we present
our theoretical evaluating procedure of the analysis. In Sec.
\ref{result} we show our study results and give relevant
discussions. Finally, a brief summary is organized  in Sec.
\ref{sum}.

\section{THEORETICAL FORMALISM}\label{theory}
\begin{figure}[htbp]
 \includegraphics*[scale=0.7]{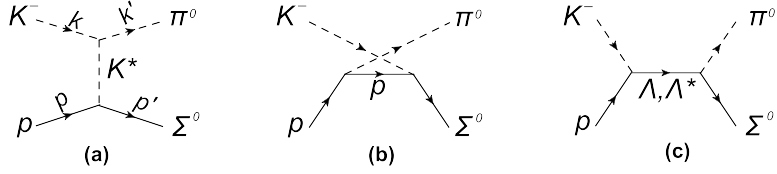}
\caption{Feynman diagrams for $K^-p\rightarrow\pi^0\Sigma^0$: (a)
t-channel $K^*$ exchange; (b) u-channel proton exchange; (c)
s-channel $\Lambda$ and its resonances exchanges.}\label{fig:graph}
\end{figure}

For the reaction $K^-p\rightarrow \pi^0\Sigma^0$, the basic
contributions come from the t-channel $K^*$ exchange, u-channel
proton exchange, s-channel $\Lambda$ and its resonances
contributions. The corresponding Feynman diagrams are shown in
Fig.~\ref{fig:graph}. In addition to the t-channel and u-channel
contributions, the s-channel contributions from five well
established 4-star $\Lambda$ and its resonances listed in
PDG~\cite{PDG} $\Lambda(1115){1\over 2}^+$, $\Lambda(1405){1\over
2}^-$, $\Lambda(1520){3\over 2}^-$, $\Lambda(1670){1\over 2}^-$ and
$\Lambda(1690){3\over 2}^-$ are always included in our analysis.

In the t-channel $K^*$ exchange process, the effective Lagrangian is
\begin{eqnarray}
   \mathcal{L}_{K^*K\pi}&=&i g_{K^*K\pi}K^*_\mu(\pi\cdot\tau\partial^\mu K-\partial^\mu \pi\cdot\tau K)\\
    {\mathcal L}_{K^*N\Sigma}&=&-g_{K^*N\Sigma}\overline{\Sigma}(\gamma_\mu K^{*\mu}-\frac{\kappa_{K^*N\Sigma}}{2M_N}\sigma_{\mu\nu}\partial^\nu K^{*\mu})N
\end{eqnarray}

The $K^*K\pi$ coupling constant can be calculated from the decay
width of $K^*\rightarrow K\pi$, getting $g_{K^*K\pi}=-3.23$. As for
the $K^*N\Sigma$ couplings, Refs.\cite{KsNLambda1,KsNLambda2} gave
two sets of values:

\begin{eqnarray*}
  g_{K^*N\Sigma}=-2.46 , ~~~~\kappa_{K^*N\Sigma}=-0.47~(NSC97a),\\
  g_{K^*N\Sigma}=-3.52 , ~~~~\kappa_{K^*N\Sigma}=-1.14~(NSC97f)
\end{eqnarray*}

Thus we limit $g_{K^*N\Sigma}$ to be between $-3.52$ and $-2.46$,
and $\kappa_{K^*N\Sigma}$ to be between $-1.14$ and $-0.47$.

The u-channel proton exchange Lagrangian is given by
\begin{eqnarray}
  {\mathcal L}_{\pi NN}&=&{g_{\pi NN}\over {2M_N}}\overline{N}\gamma^\mu\gamma^5\partial_\mu\pi\cdot\tau N\\
  {\mathcal L}_{K N\Sigma}&=&{g_{K N\Sigma}\over {M_N + M_\Sigma}} \overline{\Sigma}\cdot\tau\gamma^\mu\gamma^5 N\partial_\mu\overline{K}
\end{eqnarray}
where $g_{\pi NN}=13.45$ and $g_{KN\Sigma}=2.69$ from the SU(3)
symmetry~\cite{Xie1405}. We allow a factor between
${1\over\sqrt{2}}$ and $\sqrt{2}$ to multiply to $g_{\pi
NN}g_{KN\Sigma}$ for consideration of SU(3) symmetry breaking
effect.

For the s-channel $\Lambda$ resonances exchanges with different
$J^P$, the effective Lagrangians are
\begin{eqnarray}
{\mathcal L}_{KN\Lambda(\frac{1}{2}^+)}&=&{g_{KN\Lambda}\over {M_N +
M_\Lambda}}\partial_\mu
\overline{K}\overline{\Lambda}\gamma^\mu\gamma_5 N+H.c.\\
{\mathcal
L}_{\Lambda({1\over{2}}^+)\pi\Sigma}&=&{g_{\Lambda\pi\Sigma}\over
{M_\Lambda +
M_\Sigma}}\overline{\Sigma}\cdot\partial_\mu{\bar\pi}\gamma^\mu\gamma_5\Lambda
+H.c.
\end{eqnarray}
\begin{eqnarray}
{\mathcal L}_{KN\Lambda(\frac{1}{2}^-)}&=&-i g_{KN\Lambda({1\over{2}}^-)}\overline{K}\overline{\Lambda} N+H.c.\\
{\mathcal L}_{\Lambda({1\over{2}}^-)\pi\Sigma}&=&-i
g_{\Lambda({1\over{2}}^-)\pi\Sigma}\overline{\Lambda}\pi\cdot\Sigma
+H.c.
\end{eqnarray}
\begin{eqnarray}
{\mathcal L}_{KN\Lambda(\frac{3}{2}^+)}&=&\frac{f_{KN\Lambda({3\over{2}}^+)}}{m_K}\partial_\mu\overline{K} \overline{\Lambda}^\mu N + H.c.\\
{\mathcal
L}_{\Lambda(\frac{3}{2}^+)\Lambda\pi}&=&\frac{f_{\Lambda({3\over{2}}^+)\pi\Sigma}}{m_\pi}
\partial_\mu\overline{\pi}\cdot\overline{\Sigma}\Lambda^\mu + H.c.
\end{eqnarray}
\begin{eqnarray}
  {\mathcal L}_{KN\Lambda(\frac{3}{2}^-)} &=&\frac{f_{KN\Lambda(\frac{3}{2}^-)}}{m_K}\partial_\mu\overline{K} \overline{\Lambda}^\mu\gamma_5 N + H.c.\\
  {\mathcal L}_{\Lambda(\frac{3}{2}^-)\pi\Sigma}&=&\frac{f_{\Lambda(\frac{3}{2}^-)\pi\Sigma}}{m_\pi} \partial_\mu\pi\overline{\Lambda}^\mu\gamma_5\Sigma+H.c.
\end{eqnarray}

For $\Lambda(1115){1\over 2}^+$, the SU(3) flavor symmetry predict
$g_{KN\Lambda}=-13.98$ and $g_{\Lambda\pi\Sigma}=9.32$. Considering
SU(3) symmetry breaking effect, we multiply a tunable factor ranged
from ${1\over\sqrt{2}}$ to $\sqrt{2}$ to
$g_{KN\Lambda}g_{\Lambda\pi\Sigma}$.

For the $\Lambda(1405){1\over 2}^-$, we adopt the PDG~\cite{PDG}
estimated mass and width for it, {\sl i.e.}, 1405.1 MeV and 50MeV,
respectively. Its coupling to $\pi\Sigma$ is obtained from its decay
width to be $g_{\Lambda\pi\Sigma}=0.9$. Since $\Lambda(1405)$ is
below the $K^-p$ threshold, $g_{KN\Lambda({1\over{2}}^-)}$ cannot be
directly evaluated from the decay approach. Nevertheless, there are
many theoretical works on this parameter. Williams~\cite{1405more}
gave a upper limit for $g_{KN\Lambda(1405)}$ of $3.0$ obtained from
hadronic scattering. The work of two-pole structure for
$\Lambda(1405)$ by Oset et al.~\cite{Oset1} gives
$|g_{\overline{K}N\Lambda}|=2.1$ for the lower resonance and
$|g_{\overline{K}N\Lambda}|=2.7$ for the upper resonance. By using a
separable potential model~\cite{sep_potential},
Xie~\cite{Xie1405threshold} gave $g_{KN\Lambda(1405)}^2/4\pi =0.27$,
{\sl i.e.}, $g_{\Lambda(1405)\overline{K}N}=1.84$ at the
$\overline{K}N$ threshold. In Ref.\cite{Xie1405}, Xie et.al. gave
$g_{\Lambda(1405)\overline{K}N}=0.77$ and
$g_{\Lambda(1405)\overline{K}N}=1.51$ from two different fitting
procedures. In our analysis, we set $g_{\Lambda(1405)\overline{K}N}$
to be a free parameter.

As listed in PDG~\cite{PDG}, $\Lambda(1520){3\over 2}^-$ has very
narrow ranges of its mass, width and branching ratios to $\overline{K}N$
and $\pi\Sigma$, we fix its mass to be $1519.5MeV$ and coupling
constants $f_{KN\Lambda}=10.5$ and $f_{\Lambda\pi\Sigma}=2.12$. We
use the energy dependent width of $\Lambda(1520)$, which contains
the Blatt-Weisskopf barrier factor~\cite{centrifugal,Chung}
\begin{equation}
\Gamma(\sqrt{s})=\Gamma_0\sum\limits_{i}\left[c_i\frac{p_{B_i}^3(\sqrt{s})M_{\Lambda^*}(E_{B_i}(\sqrt{s})
-M_{B_i})
B^2_2(p_{B_i}(\sqrt{s}))}{p_{B_i}^3(M_{\Lambda^*})\sqrt{s}(E_{B_i}(M_{\Lambda^*})
-M_{B_i})
B^2_2(p_{B_i}(M_{\Lambda^*}))}\right]\diagup\sum\limits_{j}c_j
\end{equation}
where $s$ is the invariant mass of $K^-p$ system, $\Gamma_0=15.6~
MeV$, $c_i$ is the branching ratio to the i-th final state,
$c_{\overline{K}N}=0.45$, and  $c_{\pi\Sigma}=0.42$~\cite{PDG}.
$p_{B_i}(W)$ and $E_{B_i}(W)$ represent the magnitude of the three
momentum and energy of the baryon in the decayed final system,
respectively,  {\sl i.e.},
$p_{B_i}^2(W)=\frac{(W^2+M_{B_i}^2-m_{M_i}^2)^2}{4W^2}-M_{B_i}^2$
and $E_{B_i}(W)=\sqrt{p_{B_i}^2(W)+M_{B_i}^2}$.
$B_2(Q)=\sqrt{13\over{Q^4+3Q^2Q_0^2+9Q_0^4}}$ is the Blatt-Weisskopf
barrier factor~\cite{centrifugal,Chung} for $l=2$, and $Q_0$ is a
hadron ''scale" parameter as a tunable parameter ranging from 0.5 to
1.5 in our analysis.

The $\Lambda(1670){1\over 2}^-$ and $\Lambda(1690){3\over 2}^-$
coupling constants can be deduced from their relevant decays widths
to $\overline{K}N$ and $\Sigma\pi$ as listed in PDG~\cite{PDG}.
Taking into account of their uncertainties, we constrain
$g_{\Lambda(1670)\pi\Sigma}g_{KN\Lambda(1670)}$ to be in the range
of $0.04\sim 0.2$, and $f_{KN\Lambda}f_{\Lambda\pi\Sigma}$ in the
range of $2.85\sim 7.62$ in our fitting. Their masses and widths are
also tunable parameters.

At each vertex, an off-shell form factor is used. For the t-channel
$K^*$ meson exchange, we use the form factor
\begin{equation}
F_{K^*}(p_{K^*}^2)=\left(\frac{\Lambda^2-m_{K^*}^2}{\Lambda^2-p_{K^*}^2}\right)^2,
\end{equation}
where $m_{K^*}$, $p_{K^*}$ and $\Lambda$ are the mass, 4-momenta,
and cut-off parameter for the exchanged $K^*$.

For the u-channel and s-channel baryon exchanges, the off-shell form
factor is in the form
\begin{equation}
  F_B(q^2,M)=\frac{\Lambda^4}{\Lambda^4+(q^2-M^2)^2}
\end{equation}
where $M$, $q$ and $\Lambda$ stand for the mass, 4-momenta and
cut-off factor of the exchanged baryon. The cut-off parameter is
constrained between $0.8$ and $1.5$ for all channels.

The propagator for the vector meson $K^*$ exchange is
\begin{equation}
  G_{K^*}(p_{K^*})=\frac{-g^{\mu\nu}+p_{K^*}^\mu p_{K^*}^\nu/m^2_{K^*}}{p_{K^*}^2-m^2_{K^*}}
\end{equation}

For the u-channel proton exchange,
the propagator is
\begin{equation}
  G_B(q)=\frac{\not\! q +m}{q^2-m^2}.
\end{equation}

For the s-channel $\Lambda(1115)$ exchange, the expression of the propagator is
\begin{equation}
  G_B(q)=\frac{\not\! q +\sqrt{s}}{q^2-m^2}.
\end{equation}

While for other $\Lambda$ unstable resonances, the propagators are in the
Breit-Weigner forms

\begin{eqnarray}
  G_R^{1\over 2}(q)&=&\frac{\not\! q +\sqrt{s}}{q^2-M^2+iM\Gamma}\\
  G_R^{3\over 2}(q)&=&\frac{\not\! q +\sqrt{s}}{q^2-M^2+iM\Gamma}(-g^{\mu\nu}+\frac{\gamma^\mu \gamma^\nu}{3}+\frac{\gamma^\mu q^\nu-\gamma^\nu q^\mu}{3 \sqrt{s}}+\frac{2q^\mu q^\nu}{3 s})
\end{eqnarray}
where $\Gamma$ is the total width of the resonance and $s$ is the
square of the invariant mass of $K^-p$ system.

The differential cross section for ${K^- p\rightarrow\pi^0\Sigma^0}$
in the center of mass frame is
\begin{equation}
   \frac{d\sigma}{d\Omega}=\frac{1}{64 \pi^2 s}\frac{\left|\bf{k'}\right|}{\left|\bf{k}\right|}\overline{\left|\mathcal M\right|}^2,
\end{equation}
where $d\Omega=2\pi d\cos\theta$, and $\theta$ is the angle between
$K^-$ and $\pi^0$ in the center of mass frame. $\bf{k}$ and
$\bf{k'}$ represent the three-momenta of $K^-$ and $\pi^0$ in the
c.m. frame, respectively. The amplitude $\mathcal M$ and its averged
square can be expressed as
\begin{eqnarray}
\mathcal M_{\lambda,\lambda'}&=&u^{\lambda'}_{\Sigma^0}(p'){\mathcal A}u^{\lambda}_p(p)=u^{\lambda'}_{\Sigma^0}(p')\sum\limits_i{\mathcal A_i}u^{\lambda}_p(p)\\
\overline{\left|\mathcal M\right|}^2&=&{1\over
2}\sum\limits_{\lambda,\lambda'}{\mathcal
{M_{\lambda,\lambda'}M_{\lambda,\lambda'}^\dag}}={1\over
2}Tr\left[(\not\! p' +M_{\Sigma^0}){\mathcal A}(\not\! p
+M_p)\gamma^0\mathcal A^\dag\gamma^0\right]
\end{eqnarray}
where $p$ and $p'$ represent the 4-momenta of proton and $\Sigma^0$
separately, $\lambda$ and $\lambda'$ stands for the spin index of
proton and $\Sigma^0$, respectively. $\mathcal A$ is the total
amplitude despite the spin functions and $\mathcal A_i$ denotes the
i-th channel partial contribution.

The $\Sigma^0$ polarization is in the form~\cite{polarization}
\begin{equation}
  P_{\Sigma^0}=2 Im \left({\mathcal M}_{{1\over 2}{1\over 2}}{\mathcal M}^*_{-{1\over 2}{1\over 2}}\right)/\overline{\left|\mathcal M\right|}^2.
\end{equation}
%where
%\begin{equation}
%  {\mathcal I(\theta)}={1\over 2}\sum\limits_{\lambda,\lambda'}{\left|\mathcal M_{\lambda,\lambda'}\right|}^2.
%\end{equation}

\section{RESULTS AND DISCUSSIONS}\label{result}

\begin{figure}[htbp]
%\centering
\includegraphics*[width=16cm]{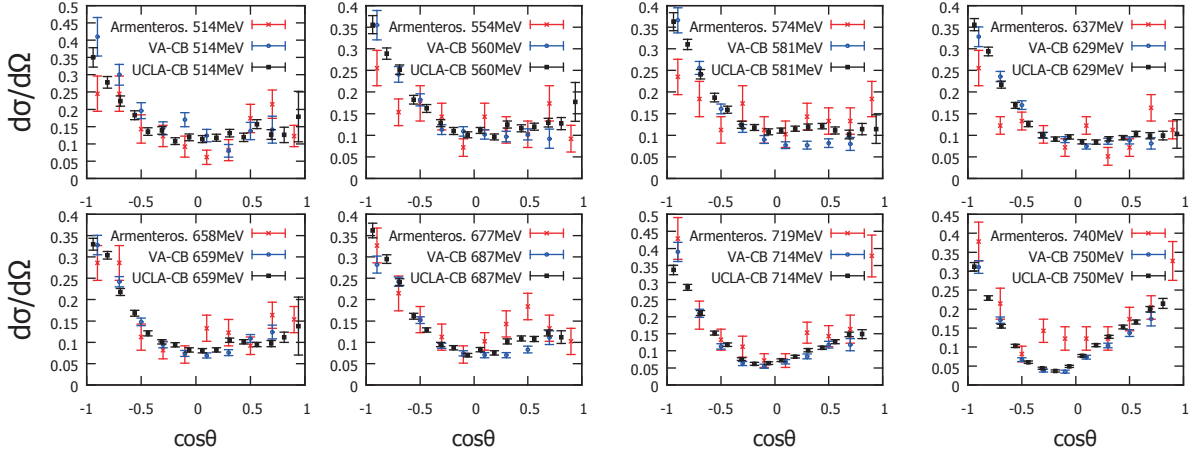}
\caption{The differential cross sections from
Ref.\cite{LowEnergyData}, the VA group~\cite{CB08} and the UCLA
group~\cite{CB09} of the Crystal Ball collaboration at similar beam
momenta.}\label{fig:3ref dcs}
\end{figure}

\begin{figure}[htbp]
%\centering
\includegraphics*[width=13cm]{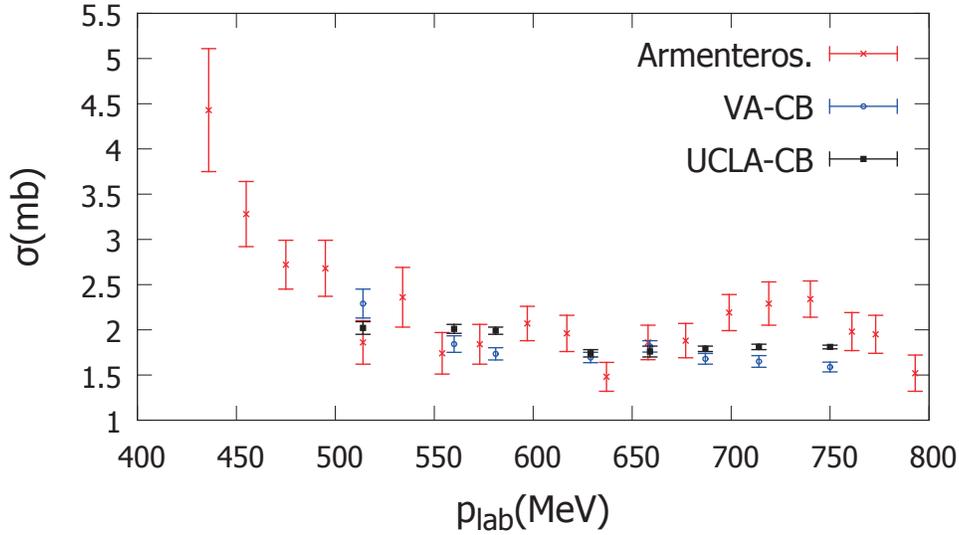}
\caption{The total cross sections from Ref.\cite{LowEnergyData}, the
VA group~\cite{CB08} and the UCLA
group~\cite{CB09}.}\label{fig:total cs}
\end{figure}

The analyzed experimental data are from Armenteros et
al.~\cite{LowEnergyData}, the VA group~\cite{CB08} and the UCLA
group~\cite{CB09} of the Crystal Ball collaboration. The
differential cross section data of these three references are shown
in Fig.~\ref{fig:3ref dcs}. We can see that the differential cross
sections from the VA group and the UCLA group of CB are compatible
with each other, while some data points from
Ref.\cite{LowEnergyData} diverge from those of the two CB groups,
but with large error bars. Fig.\ref{fig:total cs} shows the total
cross section data of the three references. The total cross sections
of the VA group and the UCLA group of CB can be smoothly extended
from the 4 lower-momentum data of Ref.\cite{LowEnergyData}. So we
will use the 4 low-momentum differential cross section data together
with those from VA group and UCLA group of CB~\cite{CB08,CB09}.

Considering the distinct polarization results of the VA group and
the UCLA group, we will first only fit the differential cross
sections given consistently by three experimental groups. Then we
will separately deal with the differential cross sections either
with the VA group polarization data or the polarization data of the
UCLA group.

Our fitting procedure is as follows. Firstly we include the
t-channel $K^*$, u-channel proton and s-channel the well-established
$\Lambda(1115)$ and its resonances $\Lambda(1405){1\over 2}^-$,
$\Lambda(1520){3\over 2}^-$, $\Lambda(1670){1\over 2}^-$ and
$\Lambda(1690){3\over 2}^-$ contributions, which contains 18 tunable
parameters, and give the results. Secondly we discuss the results by
including an additional $\Lambda$ resonance with $J^P={1\over
2}^+,~~{1\over 2}^-,~~{3\over 2}^+$ or ${3\over 2}^-$ to s-channel.
Then we try to add additional 2, 3, and 4 $\Lambda$ resonances to
see the improvement of the description of the experimental data.
Including an additional resonance increases the tunable parameters
by 4, {\sl i.e.}, the cut-off parameter, mass, width and product of
coupling constants to $\overline{K}N$ and $\pi\Sigma$ of the
resonance.

\subsection{Results of fitting only the differential cross section data from Refs.\cite{LowEnergyData,CB08,CB09}}

\begin{figure}[htbp]
%\centering
\includegraphics*[width=17cm]{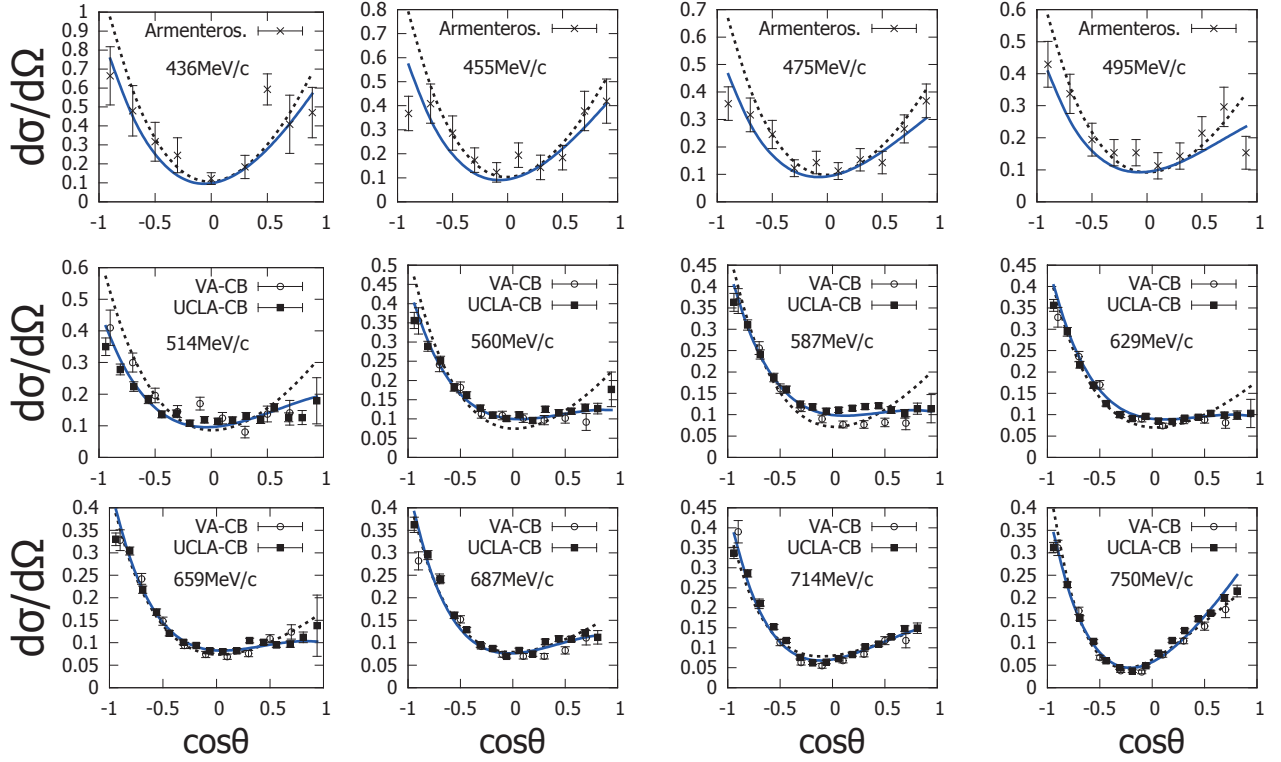}
\caption{(Color online) Fit compared with differential cross section
data from Refs.\cite{LowEnergyData,CB08,CB09}. The dashed lines show
results with inclusion of only five 4-star $\Lambda$ resonances in
s-channel; the blue solid lines represent the results of including
an additional $\Lambda({1\over 2}^+)$ resonance.}\label{fig:dcs}
\end{figure}

When only fitting the 236 differential cross section data points of
Refs.\cite{LowEnergyData,CB08,CB09}, the fit without adding
additional $\Lambda$ resonance in s-channel has $\chi^2=1565$. The
fit compared with the experimental data is shown in
Fig.~\ref{fig:dcs} by the dashed lines.

When adding one additional resonance, the best fit is to add a
$J^P={1\over 2}^+$ resonance, with mass around 1582 MeV, width about
142 MeV, leading to a $\chi^2=654$. The fitting results are shown in
Fig.~\ref{fig:dcs}. The fitted parameters and uncertainties for
$\Lambda(1670){1\over 2}^-$, $\Lambda(1690){3\over 2}^-$, and the
added $\Lambda({1\over 2}^+)$ of this solution are shown in
Table~\ref{dcs1p}. The fitted couplings for t-channel $K^*$,
u-channel proton and s-channel $\Lambda(1115)$ and $\Lambda(1405)$
are shown in Table~\ref{dcs_tu}.

\begin{table*}[htbp]
\caption{Fitted parameters of $\Lambda(1670){1\over 2}^-$,
$\Lambda(1690){3\over 2}^-$ and the additional $\Lambda({1\over
2}^+)$ for the lowest $\chi^2$ result when adding one additional
resonance.}
  \label{dcs1p}
  \centering
   \begin{tabular}{c|c|c|c|c|c|c}
   \hline\hline
   \ & mass(MeV)(PDG estimate) &
    $\Gamma_{tot}$(MeV)(PDG estimate) & $\sqrt{\Gamma_{\pi\Sigma}\Gamma_{\overline{K} N}}/\Gamma_{tot}$ (PDG range)\\
    \hline
    $\Lambda(1670)\frac{1}{2}^-$ & $1701.8\pm{3.5} $(1660,1680) & $127.9\pm{1}$(25,50) & $-0.38\pm{0.043}$(-0.38,-0.23) \\
    \hline
    $\Lambda(1690)\frac{3}{2}^-$ & $1683.8\pm{1.5} $(1685,1695) & $42.4\pm{4.8}$(50,70) & $-0.228\pm{0.037}$(-0.34,-0.25) \\
    \hline
    $\Lambda(1600)\frac{1}{2}^+$ & $1581.7\pm{32} $(1560, 1700) & $142.5\pm{4.5}$(50,250) & $-0.365\pm{0.01}$(-0.33,0.28) \\
    \hline\hline
    \end{tabular}\\
        \vspace{0.3cm}
    \caption{Fitted coupling constants for t-channel, u-channel, s-channel $\Lambda(1115)$ and $\Lambda(1405){1\over 2}^-$.  }
    \label{dcs_tu}
      \begin{tabular}{c|c|c|c|c}
         \hline\hline
          $g_{K^*N\Sigma}$ (Model)  & $g_{K^*N\Sigma}\kappa_{K^*N\Sigma}$ (Model)  & $g_{\pi NN}g_{KN\Sigma}$ [SU(3)]  & $g_{KN\Lambda}g_{\Lambda\pi\Sigma}$ [SU(3)]  & $g_{KN\Lambda^*}g_{\Lambda^*\pi\Sigma}$\\
           \hline
           $-3.52\pm{0.69}$(-3.52, -2.46)& $-1.14\pm{0.06}$(-1.14,-0.47)& $33.76\pm{1.73}$(36.18)& $92.06\pm{4.7}$(130.3)& $2.97\pm{0.15}$\\
          \hline\hline
    \end{tabular}
\end{table*}

Instead of the ${1\over 2}^+$ resonance, when adding a ${3\over
2}^-$ resonance, the $\chi^2$ is 687, with the resonance's mass
about 1526 MeV and width near 43 MeV. The other results for adding
one additional resonance are $\chi^2=1029$ for adding one
$J^P={3\over 2}^+$ resonance and $\chi^2=885$ for adding one
$J^P={1\over 2}^-$ resonance.

When adding two additional resonances, the lowest $\chi^2$ equaling
540 is given by adding a ${1\over 2}^+$ resonance and a ${3\over
2}^-$ resonance. The parameters for the ${1\over 2}^+$resonance are
mass around $1604\pm 3.3$ MeV, width about $248\pm 3.4$ MeV, and
coupling $\sqrt{\Gamma_{\pi\Sigma}\Gamma_{\overline{K}
N}}/\Gamma_{tot}=-0.31\pm 0.02$. The ${3\over 2}^-$ resonance's
mass, width and branching ratio are $1535\pm 3.3$ MeV, $29\pm 8$
MeV, and $\sqrt{\Gamma_{\pi\Sigma}\Gamma_{\overline{K}
N}}/\Gamma_{tot}=-0.11\pm 0.02$, respectively.

The second lowest result for adding two additional resonances is to
add a ${3\over 2}^+$ resonance in additional to the ${1\over 2}^+$
resonance, leading a $\chi^2$ of 546, which is very close to the
result by adding a ${3\over 2}^-$ resonance in addition to the
${1\over 2}^+$ resonance. The fitted parameters for the ${3\over
2}^+$ resonance are mass about $1680\pm 0.8$ MeV, width near $39\pm
1.3$ MeV, and branching ratio
$\sqrt{\Gamma_{\pi\Sigma}\Gamma_{\overline{K}
N}}/\Gamma_{tot}=0.11\pm 0.003$. Meanwhile, the mass, width and
couplings of ${1\over 2}^+$ are shifted to $1574\pm 0.4$ MeV,
$132\pm 0.7$ MeV, and $\sqrt{\Gamma_{\pi\Sigma}\Gamma_{\overline{K}
N}}/\Gamma_{tot}=-0.34\pm 0.001$.

The best result for adding 3 resonances is to add one ${1\over 2}^+$
resonance, one ${3\over 2}^-$ resonance, and one ${3\over 2}^+$
resonance, with the $\chi^2$ equaling 418 for the 236 experimental
data. The adjusted parameters for the established $\Lambda(1670)$
and $\Lambda(1690)$ together with the additional three resonances
are shown in Table~\ref{dcs3r}.

\begin{table*}[htbp]
\caption{Fitted parameters when additionally adding a ${1\over 2}^+$
resonance, a ${3\over 2}^-$ resonance, and a ${3\over 2}^+$
resonance for the result with $\chi^2=418$.}
  \label{dcs3r}
  \centering
  \begin{tabular}{c|c|c|c}
     \hline\hline
    \  & mass(MeV) (PDG estimate) &$\Gamma_{tot}$(MeV)(PDG estimate) & $\sqrt{\Gamma_{\pi\Sigma}\Gamma_{\overline{K} N}}/\Gamma_{tot}$ (PDG range)\\
    \hline
    $\Lambda(1670)\frac{1}{2}^-$ & $1662.6\pm{0.5} $(1660,1680) & $50\pm{18.3}$(25,50) & $-0.21\pm{0.004}$(-0.38,-0.23) \\
    \hline
    $\Lambda(1690)\frac{3}{2}^-$ & $1695\pm{28.8} $(1685,1695) & $60.3\pm{9.1}$(50,70) & $-0.051\pm{0.015}$(-0.34,-0.25) \\
    \hline
    $\Lambda(1600)\frac{1}{2}^+$ & $1574.7\pm 0.5 $ (1560, 1700) & $81.9\pm 1.1$ (50,250) & $-0.265\pm 0.002$ (-0.33,0.28) \\
    \hline
    additional $\frac{3}{2}^-$ & $1513.6\pm 0.8 $ & $230\pm 2.2$ & $-0.064\pm 0.0003$\\
    \hline
    additional $\frac{3}{2}^+$ & $1682.3\pm 0.8 $&  $132\pm 0.9 $& $0.287\pm 0.002$\\
    \hline\hline
    \end{tabular}
\end{table*}

Since adding these additional resonances significantly reduces the
contribution of $\Lambda(1690)\frac{3}{2}^-$, we examine whether it
is really needed by the data. It is found that dropping the
$\Lambda(1690)\frac{3}{2}^-$ only increases the $\chi^2$ by 1 to be
419, while dropping any other resonance will increase the $\chi^2$
by more than 120. This indicates that the
$\Lambda(1690)\frac{3}{2}^-$ is indeed not needed by the data. The
fitted parameters are shown in Table~\ref{dcs3rB}.

\begin{table*}[htbp]
  \caption{Fitted parameters with $\chi^2=419$ when dropping $\Lambda(1690)\frac{3}{2}^-$.}
  \label{dcs3rB}
  \centering
  \begin{tabular}{c|c|c|c}
     \hline\hline
    \  & mass(MeV) (PDG estimate) &
    $\Gamma_{tot}$(MeV)(PDG estimate) & $\sqrt{\Gamma_{\pi\Sigma}\Gamma_{\overline{K} N}}/\Gamma_{tot}$ (PDG range)\\
    \hline
    $\Lambda(1670)\frac{1}{2}^-$ & $1660.9\pm{0.4} $(1660,1680) & $48.3\pm{0.8}$(25,50) & $-0.22\pm{0.003}$(-0.38,-0.23) \\
    \hline
    $\Lambda(1600)\frac{1}{2}^+$ & $1576.3\pm 0.5 $ (1560, 1700) & $80.7\pm 1.1$ (50,250) & $-0.273\pm 0.002$ (-0.33,0.28) \\
    \hline
    additional $\frac{3}{2}^-$ & $1511.2\pm 1 $ & $256\pm 2.9$ & $-0.054\pm 0.003$\\
    \hline
    additional $\frac{3}{2}^+$ & $1679.8\pm 0.7 $&  $115.3\pm 0.8 $& $0.295\pm 0.002$\\
    \hline\hline
     \end{tabular}
     \\
    \vspace{0.6cm}
      \begin{tabular}{c|c|c|c|c}
               \hline\hline
          $g_{K^*N\Sigma}$ (Model)  & $g_{K^*N\Sigma}\kappa_{K^*N\Sigma}$ (Model)  & $g_{\pi NN}g_{KN\Sigma}$ [SU(3)]  & $g_{KN\Lambda}g_{\Lambda\pi\Sigma}$ [SU(3)]  & $g_{KN\Lambda^*}g_{\Lambda^*\pi\Sigma}$\\
           \hline
           $-2.46\pm{1.06}$(-3.52, -2.46)& $-0.52\pm{0.37}$(-1.14,-0.47)& $51.2\pm{24.1}$(36.18)& $92.1\pm{59.6}$(130.3)& $2.49\pm{0.02}$\\
           \hline\hline
      \end{tabular}
\end{table*}

Further, when we replace the ${3\over 2}^+$ resonance by the well
established $\Lambda(1890){3\over 2}^+$ with mass from 1850 to 1910
MeV and width from 60 to 200 MeV ~\cite{PDG}, the $\chi^2$ is 512.
The fitted parameters of the five resonances are shown in
Table~\ref{dcs3rC}. This demonstrates that the new ${3\over 2}^+$
resonance around 1680 MeV cannot be replaced by the well established
$\Lambda(1890){3\over 2}^+$.

\begin{table*}[htbp]
  \caption{Fitted resonance parameters with $\chi^2=512$ when replacing the new $3/2^+$ resonance by $\Lambda(1890)\frac{3}{2}^+$.}
  \label{dcs3rC}
  \centering
  \begin{tabular}{c|c|c|c}
     \hline\hline
    \ & mass(MeV)(PDG estimate) &
    $\Gamma_{tot}$(MeV)(PDG estimate) & $\sqrt{\Gamma_{\pi\Sigma}\Gamma_{\overline{K} N}}/\Gamma_{tot}$ (PDG range)\\
    \hline
    $\Lambda(1670)\frac{1}{2}^-$ & $1670.9\pm{0.5} $(1660,1680) & $50\pm{1.2}$ (25,50) & $-0.19\pm{0.006}$(-0.38,-0.23) \\
      \hline
      $\Lambda(1690)\frac{3}{2}^-$ & $1695\pm{1.6} $(1685,1695) & $70\pm{13.1}$ (50,70) & $-0.165\pm{0.005}$(-0.34,-0.25) \\
     \hline
      $\Lambda(1600)\frac{1}{2}^+$ & $1563.2\pm 0.2 $(1560, 1700) & $159\pm 0.4$(50,250) & $-0.337\pm 0.001$(-0.33,0.28) \\
    \hline
    $\Lambda(1890)\frac{3}{2}^+$ & $1850\pm 0.5 $ (1850,1910) & $200\pm 0.1$(60,200) &$ 0.99\pm 0.002$\\
    \hline
    additional $\frac{3}{2}^-$ & $1558.3\pm 0.9 $ & $130.6\pm 2.8$ & $-0.05\pm 0.0002$ \\
      \hline
      \end{tabular}
\end{table*}

 \begin{table*}[htbp]
  \caption{Fitted parameters for the best fit of $\chi^2=403$ when adding 4 additional resonances.}
  \label{dcs4r}
  \centering
  \begin{tabular}{c|c|c|c}
     \hline\hline
    \ $J^P$ & mass(MeV) &
    $\Gamma_{tot}$(MeV) & $\sqrt{\Gamma_{\pi\Sigma}\Gamma_{\overline{K} N}}/\Gamma_{tot}$ \\
    \hline
    $\frac{1}{2}^+$ & $1580.3\pm 1.3 $ & $67.8\pm2.6 $& $-0.24\pm0.002$ \\
    \hline
    $\frac{1}{2}^+$ & $1544.8\pm 1$ & $36.3\pm 2.5$ &$ -0.21\pm 0.006$ \\
    \hline
    $\frac{3}{2}^-$ & $1505.2\pm 1.4$ & $274.4\pm 2.2$ & $-0.049\pm0.0002$\\
    \hline
    $\frac{3}{2}^+$ & $1680.7\pm 1.1$ &  $144.9\pm 2.3$ & $0.281\pm 0.002$\\
    \hline\hline
    \end{tabular}
\end{table*}

\begin{figure}[htbp]
{\includegraphics*[width=15cm]{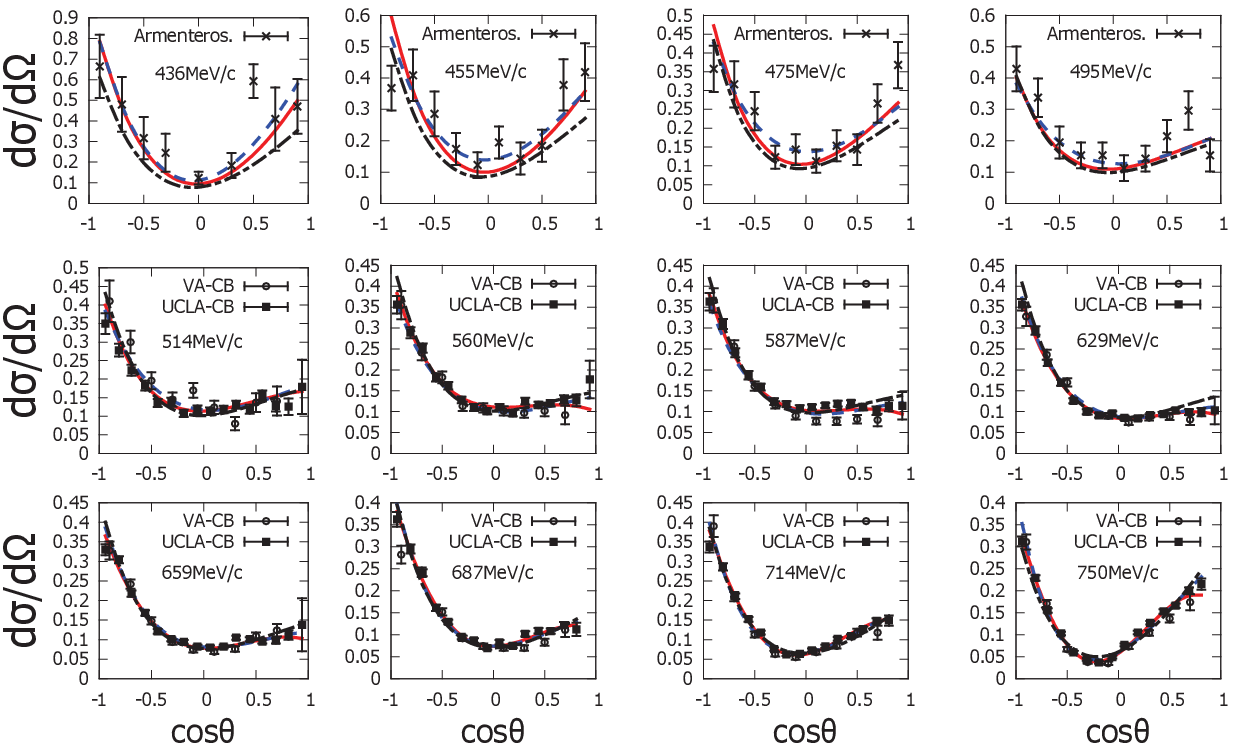}}
{\includegraphics*[width=15cm]{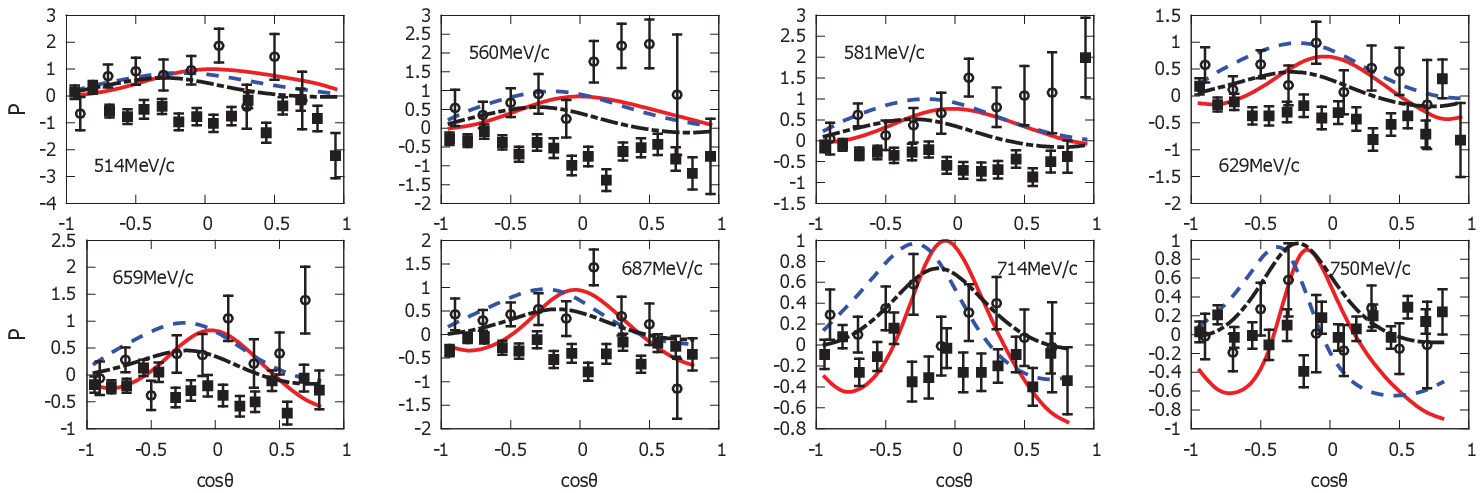}} \caption{(Color
online) The most favored fit (red solid lines) to differential
cross section data from Refs.\cite{LowEnergyData,CB08,CB09}, and
corresponding prediction to the polarizations compared with data of
Refs.\cite{CB08,CB09}.  As comparison, fits by dropping the
$\frac{3}{2}^+$ resonance or replacing it by
$\Lambda(1690)\frac{3}{2}^-$ are shown with blue dashed and black
dot-dashed lines, respectively. }\label{fig:dcs_3r}
\end{figure}

The best fit by adding 4 additional resonances has $\chi^2=403$, by
adding two ${1\over 2}^+$ resonances, one ${3\over 2}^-$ resonance
and one ${3\over 2}^+$ resonance. Their fitted parameters are shown
in Table~\ref{dcs4r}. The two ${1\over 2}^+$ resonances strongly
overlap and can be regarded as some modification to the shape of one
resonance. Moreover, the fit improves $\chi^2$ only by 15 with 4
additional parameters. This suggests that no evidence for any more
resonances from the data.

From our above investigation, we regard the fit given in
Table~\ref{dcs3rB} as our most favored fit to the CB data on the
differential cross sections. In this most favored fit, the PDG
4-star resonance $\Lambda(1670)\frac{1}{2}^-$ and 3-star resonance
$\Lambda(1600)\frac{1}{2}^+$  are definitely needed with fitted
parameters compatible with PDG values; the PDG 4-star resonance
$\Lambda(1690)\frac{3}{2}^-$ is dropped, replaced by a new
$\Lambda(1680)\frac{3}{2}^+$ resonance; an additional broad $3/2^-$
contribution couples to this channel weakly and may be regarded as
modification to the tail of $\Lambda(1520)\frac{3}{2}^-$. The fitted
results of this most favored solution are shown in
Fig.~\ref{fig:dcs_3r}, together with the predicted polarizations of
this solution compared with two sets of CB polarization
data~Refs.\cite{CB08,CB09}. The predicted polarizations are more
inclined to the data by VA group~\cite{CB08}. As comparison, fits by
dropping the $\frac{3}{2}^+$ resonance with $\chi^2=763$ or
replacing it by $\Lambda(1690)\frac{3}{2}^-$ with $\chi^2=540$ are
also shown in Fig.~\ref{fig:dcs_3r} by blue dashed and black
dot-dashed lines, respectively.

Since the two sets of  CB polarization data~Refs.\cite{CB08,CB09}
are not consistent with each other, we will examine how each set of
the polarization data influences our solution separately in the
following two subsections.

\subsection{Fitting the differential cross sections of Refs.~\cite{LowEnergyData,CB08,CB09} and polarization data from the VA group of CB~\cite{CB08} }

Based on our most favored solution in last subsection, we refit the
data by including the polarization data from the VA group of
CB~\cite{CB08}. The $\chi^2$ is 550 for the 308 experimental data.
The refitted parameters of $\Lambda(1670)\frac{1}{2}^-$ and the
three additional resonances as well as the couplings for t-channel
$K^*$, u-channel proton and s-channel $\Lambda(1115)$ and
$\Lambda(1405)$ are shown in Table~\ref{pCB083rB}, The fits compared
with data are shown in Fig.~\ref{fig:pCB08}.

\begin{table*}[htbp]
  \caption{Refitted parameters for our most favored solution with $\chi^2=550$ when including polarization data from the VA group of CB~\cite{CB08}.}
  \label{pCB083rB}
  \centering
  \begin{tabular}{c|c|c|c}
     \hline\hline
    \ & mass(MeV)(PDG estimate) &
    $\Gamma_{tot}$(MeV)(PDG estimate) & $\sqrt{\Gamma_{\pi\Sigma}\Gamma_{\overline{K} N}}/\Gamma_{tot}$ (PDG range)\\
    \hline
    $\Lambda(1670)\frac{1}{2}^-$ & $1662.5\pm{0.3} $(1660,1680) & $50\pm{0.7}$ (25,50) & $-0.29\pm{0.003}$(-0.38,-0.23) \\
    \hline
    $\Lambda(1600)\frac{1}{2}^+$ & $1575.2\pm 0.6 $(1560, 1700) & $94.8\pm 1$(50,250) & $-0.293\pm 0.002$(-0.33,0.28) \\
    \hline
    additional $\frac{3}{2}^-$ & $1506.9\pm 1.4 $ & $334.4\pm 3.4$ & $-0.04\pm 0.002$ \\
    \hline
    additional $\frac{3}{2}^+$ & $1687.7\pm 1 $ & $112.7\pm 0.8$ &$ 0.297\pm 0.002$\\
    \hline
      \end{tabular}
   \vspace{0.6cm}
%    \caption{Refitted coupling constants for t-channel, u-channel and s-channel $\Lambda(1115)$ and $\Lambda(1405){1\over 2}^-$.  }
%    \label{PCB08_tu}
      \begin{tabular}{c|c|c|c|c}
         \hline\hline
         $g_{K^*N\Sigma}$ (Model)  & $g_{K^*N\Sigma}\kappa_{K^*N\Sigma}$ (Model)  & $g_{\pi NN}g_{KN\Sigma}$ [SU(3)]  & $g_{KN\Lambda}g_{\Lambda\pi\Sigma}$ [SU(3)]  & $g_{KN\Lambda^*}g_{\Lambda^*\pi\Sigma}$\\
         \hline
         $-3.52\pm{0.75}$(-3.52, -2.46)& $-1.14\pm{0.11}$(-1.14,-0.47)& $28.8\pm{1.3}$(36.18)& $92.1\pm{9.4}$(130.3)& $2.16\pm{0.03}$\\
         \hline\hline
    \end{tabular}
\end{table*}

\begin{figure}[htbp]\label{fig:pCB08}
{\includegraphics*[width=15cm]{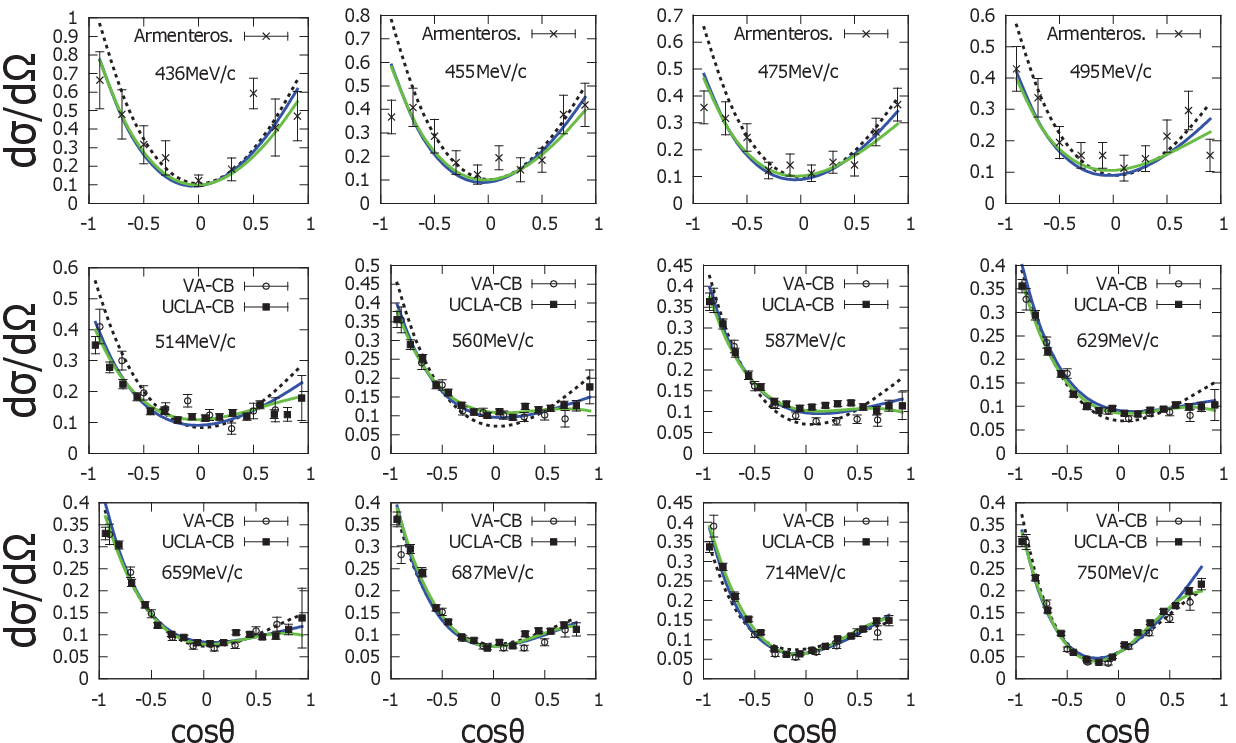}}
{\includegraphics*[width=15cm]{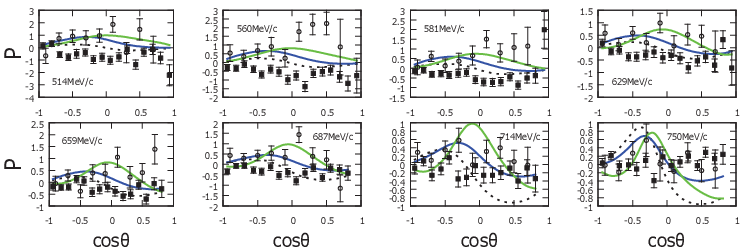}} \caption{(Color online)
Fits compared with differential cross data of
Refs.~\cite{LowEnergyData,CB08,CB09} and polarization data of
Refs.~\cite{CB08,CB09}. The dashed lines represent the result with
4-star resonances only, the blue solid lines stand for the result
when adding the 3-star $\Lambda(1600){1\over 2}^+$ resonance, and
the green solid lines show the result of the most favored solution.}\label{fig:pCB08}
\end{figure}

The refitted parameters are quite similar with those without
including the polarization data. Once again, the 4-star resonance
$\Lambda(1690)\frac{3}{2}^-$ is not needed. Adding it into our
present solution only improves $\chi^2$ by 0.5. Improvement by
adding any new resonance with other quantum numbers is also
insignificant. Dropping either $3/2^-$ or $3/2^+$ resonance in
Table~\ref{pCB083rB} will increase the $\chi^2$ by more than 100.

\subsection{Fitting the differential cross sections of Refs.~\cite{LowEnergyData,CB08,CB09} and polarization data from the UCLA group of CB~\cite{CB09} }

\begin{figure}[htbp]
{\includegraphics*[width=15cm]{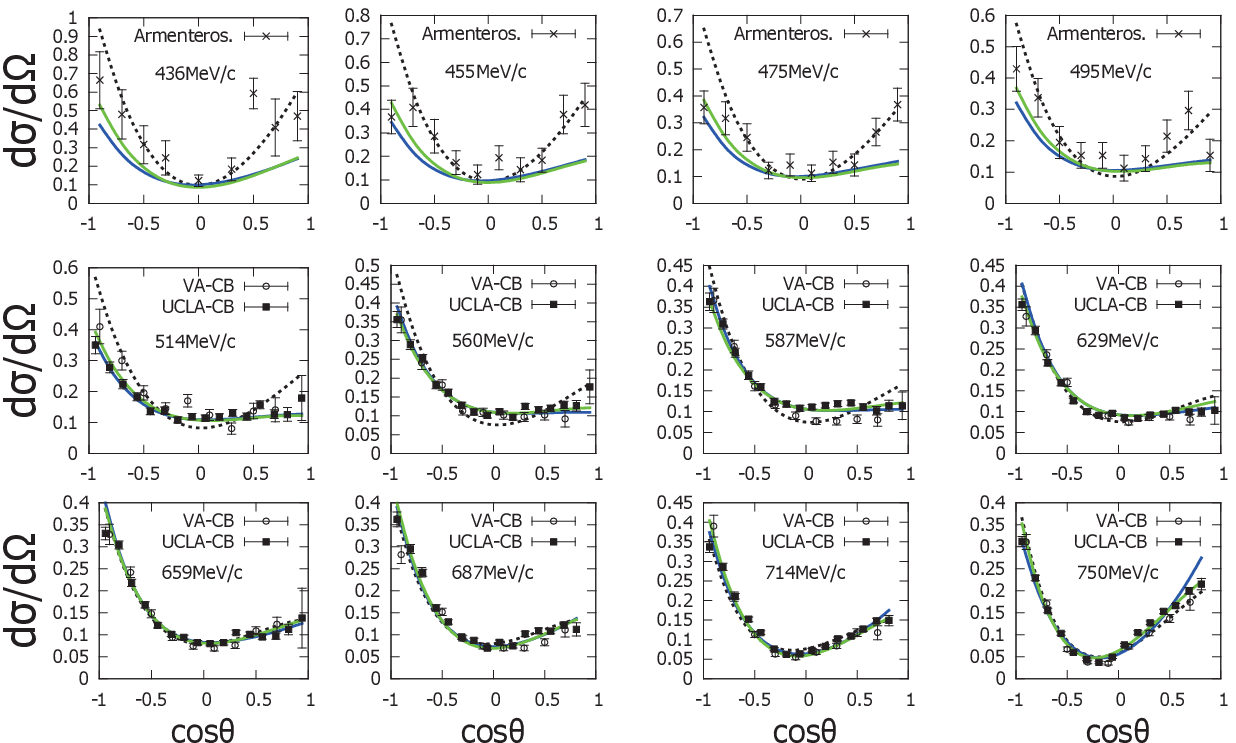}}
{\includegraphics*[width=15cm]{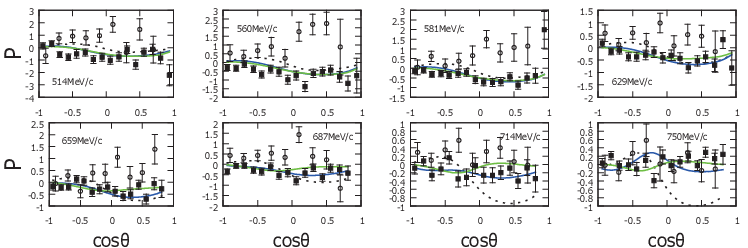}} \caption{(Color online)
Fits compared with differential cross data of
Refs.~\cite{LowEnergyData,CB08,CB09} and polarization data of
Refs.~\cite{CB08,CB09}. The dashed lines represent the result with
4-star resonances only, the blue solid lines stand for the result
when adding the 3-star $\Lambda(1600){1\over 2}^+$ resonance, and
the green solid lines show the result of our most favored solution.}\label{fig:pCB09}
\end{figure}

If we refit the data using the polarization data from the UCLA
group~\cite{CB09} instead of the VA group of CB~\cite{CB08} for our
most favored solution, the $\chi^2$ is 881 for the 360 experimental
data points. The refitted parameters of $\Lambda(1670)\frac{1}{2}^-$
and the three additional resonances  as well as  the couplings for
t-channel $K^*$, u-channel proton and s-channel $\Lambda(1115)$ and
$\Lambda(1405)$  are shown in Table~\ref{pCB093rB}. The fits
compared with data are shown in Fig.~\ref{fig:pCB09}.

\begin{table*}[htbp]
  \caption{Refitted parameters for our most favored solution with $\chi^2=881$ when when
including polarization data from the UCLA group of CB~\cite{CB09}.}
  \label{pCB093rB}
  \centering
  \begin{tabular}{c|c|c|c}
     \hline\hline
    \ & mass(MeV)(PDG estimate) &
    $\Gamma_{tot}$(MeV)(PDG estimate) & $\sqrt{\Gamma_{\pi\Sigma}\Gamma_{\overline{K} N}}/\Gamma_{tot}$ (PDG range)\\
    \hline
    $\Lambda(1670)\frac{1}{2}^-$ & $1674.2\pm{0.6} $(1660,1680) & $30\pm{1}$ (25,50) & $-0.12\pm{0.004}$(-0.38,-0.23) \\
        \hline
    $\Lambda(1600)\frac{1}{2}^+$ & $1557.1\pm 0.4 $(1560, 1700) & $169.7\pm 0.7$(50,250) & $-0.36\pm 0.001$(-0.33,0.28) \\
    \hline
    additional $\frac{3}{2}^-$ & $1585.4\pm 2.4 $ & $58.4\pm 4.5$ & $-0.035\pm 0.001$ \\
    \hline
    additional $\frac{3}{2}^+$ & $1665.6\pm 1.1 $ & $136.5\pm 3$ &$ 0.136\pm 0.003$\\
    \hline
      \end{tabular}
      \vspace{0.9cm}
%    \caption{Refitted coupling constants for t-channel, u-channel and s-channel $\Lambda(1115)$ and $\Lambda(1405){1\over 2}^-$.  }
%    \label{PCB09_tu}
      \begin{tabular}{c|c|c|c|c}
         \hline\hline
         $g_{K^*N\Sigma}$ (Model)  & $\kappa_{K^*N\Sigma}$ (Model)  & $g_{\pi NN}g_{KN\Sigma}$ [SU(3)]  & $g_{KN\Lambda}g_{\Lambda\pi\Sigma}$ [SU(3)]  & $g_{KN\Lambda^*}g_{\Lambda^*\pi\Sigma}$\\
         \hline
         $-3.47\pm{0.8}$(-3.52, -2.46)& $-0.92\pm{0.5}$(-1.14,-0.47)& $39.13\pm{0.5}$(36.18)& $92.1\pm{4.6}$(130.3)& $0.5\pm{0.06}$\\
         \hline\hline
    \end{tabular}
\end{table*}

Compared with using the polarization data of VA group, the refitted
parameters by using the UCLA data have larger difference from those
without including the polarization data. Including the 4-star
resonance $\Lambda(1690)\frac{3}{2}^-$ improves $\chi^2$ by 22,
still much less significant than other resonances. Dropping the
$3/2^+$ or $3/2^-$ resonance in Table~\ref{pCB093rB} will increase
the $\chi^2$ by 128 or 68, respectively.

\section{SUMMARY}\label{sum}

We have analyzed the $K^-p\to\pi^0\Sigma^0$ reaction using an
effective Lagrangian approach. By fitting different sets of
experimental data by CB Collaboration, we obtain the following
conclusions.

The 4-star $\Lambda(1670){1\over 2}^-$ and 3-star
$\Lambda(1600){1\over 2}^+$ resonances listed in PDG~\cite{PDG} are
definitely needed no matter which set of CB data is used. As shown
in Table~\ref{dcs3rB} for our most favored solution, the fitted
parameters for these two resonances are consistent with their PDG
values. In addition, there is strong evidence for the existence of a
new $\Lambda({3\over 2}^+)$ resonance around 1680 MeV. It improves
$\chi^2$ by more than 100 no matter which set of data is used. It
gives large contribution to this reaction, replacing the
contribution from the 4-star $\Lambda(1690){3\over 2}^-$ resonance
included by previous fits to this reaction. Including some broad
$3/2^-$ contribution also improves the $\chi^2$ significantly. It
couples to this channel weakly and may be regarded as modification
to the tail of $\Lambda(1520)\frac{3}{2}^-$.

Replacing the PDG $\Lambda(1690){3\over 2}^-$ resonance by a new
$\Lambda(1680){3\over 2}^+$ resonance has important implications on
hyperon spectroscopy and its underlying dynamics. While the
classical qqq constituent quark model~\cite{Capstick1986} predicts
the lowest $\Lambda({3\over 2}^+)$ resonance to be around 1900 MeV
in consistent with the $\Lambda(1890){3\over 2}^+$ listed in PDG,
the penta-quark dynamics~\cite{pentq1} predicts to be below 1700 MeV
in consistent with $\Lambda(1680){3\over 2}^+$ claimed in this work.

A recent analysis~\cite{Xie_eta} of CB data on the
$K^-p\to\eta\Lambda$ reaction requires a $\Lambda({3\over 2}^-)$
resonance with mass about 1670 MeV and width about 1.5 MeV instead
of the well established $\Lambda(1690){3\over 2}^-$ resonance with
width around 60 MeV. Together with $N^*(1520){3\over 2}^-$,
$\Sigma(1542){3\over 2}^-$ suggested in Ref.~\cite{Sigma} and either
$\Xi(1620)$ or $\Xi(1690)$, they fit in a nice $3/2^-$ baryon nonet
with large penta-quark configuration, {\sl i.e.}, $N^*(1520)$ as
$|[ud]\{uq\}\bar q>$ state, $\Lambda(1520)$ as $|[ud]\{sq\}\bar q>$
state, $\Lambda(1670)$ as $|[ud]\{ss\}\bar s>$ state, and
$\Xi(16xx)$ as $|[ud]\{ss\}\bar q>$ state. Here $\{q_1q_2\}$ means a
diquark with configuration of flavor representation ${\bf 6}$, spin
1 and color $\bar 3$. The $\Lambda(1670)$ as $|[ud]\{ss\}\bar s>$
state gives a natural explanation for its dominant $\eta\Lambda$
decay mode with a very narrow width due to its very small phase
space meanwhile a D-wave decay~\cite{ZouHyp}.

It would be very important to re-check other relevant reactions
whether the new claimed $\Lambda(1680){3\over 2}^+$ is also needed
there and may replace the PDG well established $\Lambda(1690){3\over
2}^-$. Further precise polarization data for KN reactions would be
very helpful to clarify the ambiguities in the determination of
spin-parities of these hyperon resonances.

\begin{acknowledgments}
Helpful discussions with Bo-Chao Liu, Pu-Ze Gao, Jia-Jun Wu, Ju-Jun
Xie, Xu Cao and W. Briscoe are gratefully acknowledged. This work is
supported by the National Natural Science Foundation of China under
Grants 11035006, 11121092, 11261130311 (CRC110 by DFG and NSFC), the
Chinese Academy of Sciences under Project No.KJCX2-EW-N01.
\end{acknowledgments}

\newcommand{\etal}{{\em et al.}}
\bibliographystyle{unsrt}

\end{document}